\documentclass{sig-alternate}

\usepackage[ruled,vlined]{algorithm2e}
\usepackage{graphicx}
\usepackage{caption}
\usepackage{subcaption}
\usepackage{multirow}
\usepackage{url}
\usepackage{hyperref}

\DeclareCaptionType{copyrightbox}

\begin{document}

\title{Timely crawling of high-quality ephemeral new content}
%
%
%
%
%


\numberofauthors{4} 
%
\author{
%
%
\alignauthor
Damien Lefortier \\ 
       \affaddr{Yandex}\\
       \email{damien@yandex-team.ru}
\alignauthor
Liudmila Ostroumova \\
       \affaddr{Yandex}\\
       \affaddr{Moscow State University}\\
       \email{ostroumova-la@yandex-team.ru}
\alignauthor
Egor Samosvat\\
       \affaddr{Yandex}\\
       \affaddr{Moscow Institute of Physics and Technology}\\
       \email{sameg@yandex-team.ru}
\and  
\alignauthor
Pavel Serdukov\\
       \affaddr{Yandex}\\
      \email{pavser@yandex-team.ru}
}
\additionalauthors{}
\date{}

\maketitle

\footnotetext[1]{The authors are given in alphabetical order}

\begin{abstract}
  Nowadays, more and more people use the Web as
  their primary source of up-to-date information. In this context,
  fast crawling and indexing of newly created Web pages has become
  crucial for search engines, especially because user traffic to a
  significant fraction of these new pages (like news, blog and forum posts)
  grows really quickly right  after they appear, but lasts only for several days.

  In this paper, we study the problem of timely finding and crawling
  of such \textit{ephemeral} new pages (in terms of user interest).
  Traditional crawling policies do not give any particular priority to
  such pages and may thus crawl them not quickly enough, and even
  crawl already obsolete content. We thus propose a new metric, well
  thought out for this task, which takes into account the decrease of
  user interest for ephemeral pages over time.

  We show that most ephemeral new pages can be found at a relatively
  small set of content sources and present a procedure for finding
  such a set.  Our idea is to periodically recrawl content sources
  and crawl newly created  pages linked from them, focusing on high-quality (in terms of user interest)
  content.   One of the main difficulties here is to divide resources between
  these two activities in an efficient way.
  We find the adaptive balance between crawls and recrawls by maximizing the proposed metric.
  Further, we incorporate search engine click logs to give our crawler an insight about the current user demands. Efficiency of our approach is finally demonstrated experimentally on real-world data.
\end{abstract}


\section{Introduction}\label{Intro}

A web crawler traditionally fulfills two purposes: discovering new
pages and refreshing already discovered pages. Both of these problems
have been extensively investigated over the past decade (see the survey paper by Olston and Najork \cite{WebCrawling}). However, recently, the role of the Web as
a media source became increasingly important as more and more people start to
use it as their primary source of up-to-date information.
This evolution forces crawlers of Web search engines to continuously
collect newly created pages as fast as possible, especially
high-quality ones.

\begin{figure}
\begin{center}
\includegraphics[width = 0.48\textwidth]{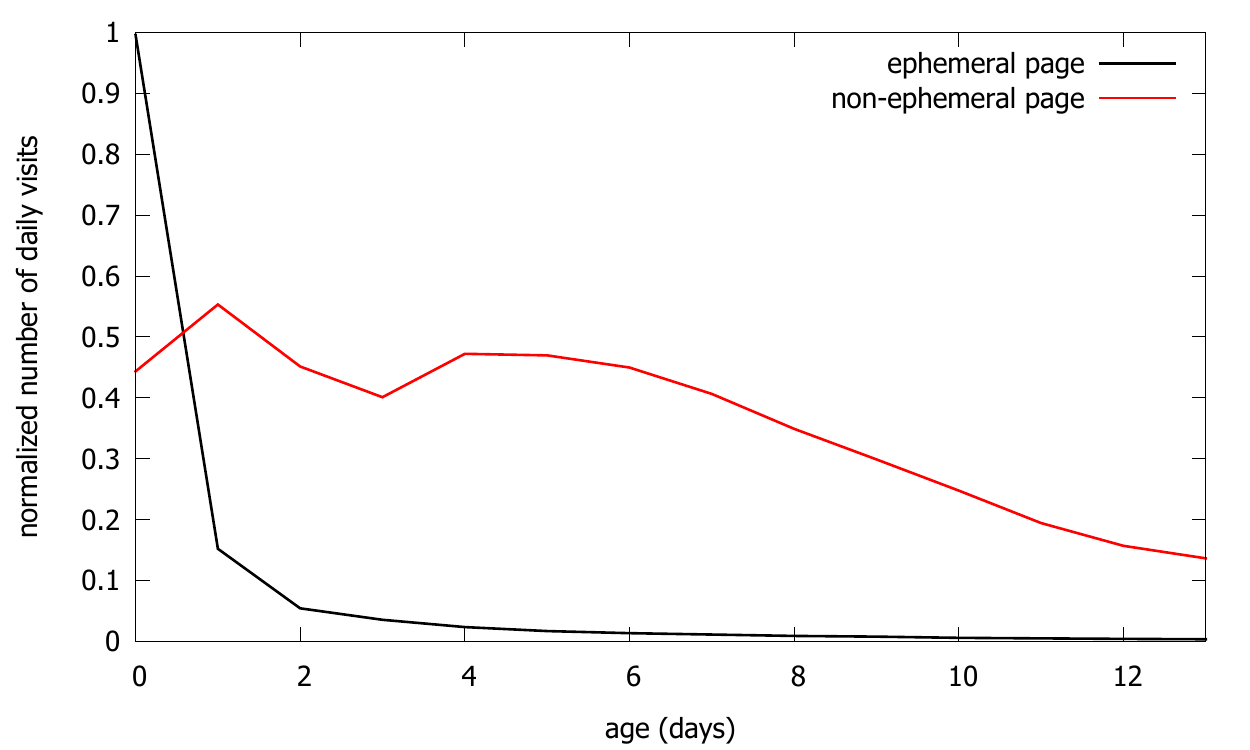}
\end{center}
\caption{Typical user interest patterns for ephemeral and
  non-ephemeral new pages}
\label{fig:yabar}
\end{figure}

Surprisingly, user traffic to many of these newly created pages grows
really quickly right after they appear, but lasts only for a few
days. For example, it was discussed in several papers that the
popularity of news decreases exponentially with time
\cite{Goyal,Moon}.  This observation naturally leads to distinguishing
two types of new pages appearing on the Web: \textit{ephemeral} and
\textit{non-ephemeral} pages. Note that here we do not consider the ephemeral content, which might be removed before it hits the index as in
\cite{RecrawlScheduling} (e.g. advertisements or the ``quote of the
day''), but we consider persistent content that is ephemeral in terms of user
interest (e.g. news, blog and forum posts).
We clustered user interest patterns of some new pages discovered in one week (see Section
\ref{ContentSources} for details), and Figure~\ref{fig:yabar} shows
the centroids of the obtained clusters.
In Section \ref{ContentSources}, we show that a significant fraction
of new pages appearing on the Web every day are ephemeral pages.

The cost of the time delay between the appearance
of such ephemeral new pages and their crawl is thus very high in terms of search engine user
satisfaction.  Moreover, if a crawler fails to find such a page during
its period of peak interest, then there might be no need to crawl it at
all.
It was reported in \cite{Recency}, that 1-2\% of user queries are extremely recency
sensitive, while even more are also recency sensitive to some extent. The problem of timely finding and crawling
ephemeral new pages is thus important, but, to the best of our
knowledge, is not studied in the literature.

Indeed, different metrics were suggested to measure the coverage and
freshness of the crawled corpus \cite{Cho1,Cho,WebCrawling}, but they
do not take into account the degradation of the profit to a
search engine contributed by these pages. Crawling policies based on such metrics may
then crawl such new pages not quickly enough, and even crawl already
obsolete content. Thus, we  need a new quality metric, well thought out for this task,
and a crawling algorithm optimized to maximize this metric over time,
that takes into account this degradation of pages' utility.

Our daily experience of using the Web also suggests that such ephemeral new
pages can be found from a relatively small set of ``hubs'' or
\emph{content sources}. We investigate this intuition and show that it is
possible and practical to find such sources at scale. Examples of
content sources are main pages of blogs, news sites, category pages of
such news sites (e.g.\ politics, economy), RSS feeds, sitemaps
\cite{Sitemaps}, etc., and one needs to periodically recrawl such
sources in order to find and crawl ephemeral new pages way before their
peak of user interest. However, frequent recrawling of all these sources and all new pages found on them requires a huge amount of
resources and is quite inefficient.
In order to solve this problem efficiently, we analyze the problem of
dividing limited resources between different tasks (coined as
\textit{holistic crawl ordering} by Oslton and Najork in
\cite{WebCrawling}), i.e., here between the task of crawling ephemeral new pages and the task of
recrawling content sources in order to discover those new pages. A possible solution for this problem is to
give a fixed quota to each policy (see, e.g., \cite{Autho}),
but we will show that such solutions based on fixed quotas are far from being optimal.

In this paper, we propose a new algorithm that dynamically
estimates, for each content source, the rate of new links appearance
in order to find and crawl newly created pages as they
appear. As a matter of fact, it is next to impossible to crawl all these new pages immediately due to resource constraints, therefore, a reasonable crawling policy
has to crawl the highest quality pages in priority.

The quality of a page can be measured in different ways, and it can, for example,
be based on the link structure of the Web graph (e.g., in-degree
\cite{Autho} or PageRank \cite{RankMass,OPIC}), or on some external
signals (e.g., query log \cite{fetterly,Olston2,Olston1} or the number
of times a page was shown in the results of a search engine
\cite{Olston2}). In this paper, we propose to use the number of clicks in order
to estimate the quality of pages, and predict the quality of
newly created pages by using
the quality of pages previously linked from each content source.
By the number of clicks, we mean the number of times a user
clicked on a link to this page on a search engine results page (SERP),
which most reliably indicates a certain level of user interest in the page's
content.  In this way, we are able, in fact, to incorporate
user feedback into the process of crawling
for our algorithm to find and crawl the best new pages.

To sum up, this paper makes the following contributions:

\begin{itemize}
\item{We formalize the problem of timely crawling of high-quality ephemeral new Web content by
    suggesting to optimize a new quality metric, which measures the
    ability of a crawing algorithm to solve this specific problem (Section
    \ref{Formalization}).}
\item{We show that most of such ephemeral new content can be found
    at a small set of content sources, and we propose a method to find
    such sources (Section \ref{ContentSources}).}
\item{We propose a practical algorithm, which periodically recrawls content
    sources and crawls newly created pages linked from them, as a
    solution of this problem.  This algorithm uses user feedback to
    estimate the quality of content sources (Section \ref{Algo}).}
\item{We validate our algorithm by comparing it to other crawling strategies on real-world data (Section~\ref{Exp}).}
\end{itemize}

Besides, in Section \ref{RelatedWork}, we review related work, while
in Section \ref{Conclusion}, we conclude the paper and discuss
possible directions for future research.

\section{Formalization of the problem}\label{Formalization}
In this section, we formalize the problem under consideration by introducing
an appropriate quality metric, which measures the
ability of a crawling algorithm to solve this problem.  As we
discussed in the introduction, we deal with pages for which user
interest grows within hours after they appear, but lasts only for
several days. The profit of crawling such ephemeral new pages thus
decreases dramatically with time.

Assume that for each page $i$, we know a decreasing function $P_i(\Delta t)$,
which is the profit of crawling this page with delay $\Delta t$ seconds
after its creation time $t_i$ (by profit, one can mean the expected number of clicks or shows on SERP).
If, finally, each page $i$ was crawled with a delay $\Delta t_i$,
we can define  the \emph{dynamic quality} of a crawler as:
\begin{equation}\label{QT}
Q_T(t) = \frac{1}{T} \sum_{i:t_i+\Delta t_i \in [t - T, t]}  P_i (\Delta t_i).
\end{equation}
In other words, the dynamic quality is the average profit gained by a crawler per second in a time window of size $T$.


The dynamic quality defined above can be useful to understand the influence of daily
and weekly trends on the performance of a crawler. Let us now define
 the \emph{overall quality} of a crawler, which allows to easily
compare different algorithms over larger time windows.  It is natural
to expect that if $T$ is large enough then the influence of season and weekly
trends of user interest will be reduced. In other words, the function $Q_T(t)$ tends to a
constant while $T$ increases. Thus, we can consider the \emph{overall quality}:
\begin{equation}\label{Q}
Q = \lim_{T\rightarrow \infty} \frac{1}{T} \sum_{i:t_i+\Delta t_i \in [0, T]}  P_i (\vartriangle t_i),
\end{equation}
which does not depend on $t$ and $T$.

In this paper, by $P_i(\Delta t)$ profit of crawling a page $i$ at time ${t_i + \Delta t}$,
we mean the total number of clicks this page will get
on a SERP after this time (ignoring any indexing delay).
In this way, we can approximate the relevance
of a page to current interests of users. From a crawler's
perspective, it is thus an appropriate measure of the contribution of
new pages to a search engine performance (given a specific ranking
method).

Alternatively, instead of the number of clicks, we could use the
number of shows, i.e., the number of times a page was shown in the top
$k$ results of a search engine. This value also reflects a crawler's
performance in the sense that we only want to crawl ephemeral new
pages, which are going to be shown to users. But, as we discuss further in this section, the
number of clicks and the number shows behave similarly and we thus use
clicks since they reflect the actual preference of users.

At this point, we have defined a new metric to measure the quality of crawling ephemeral new pages.
We are going to use this metric to
validate our algorithm in Section~\ref{Exp}.  Note, that this metric
can only be evaluated with some delay because we need to wait when
crawled pages are not shown anymore to take their profit, i.e., their
number of clicks, into account.



\begin{figure*}
        \centering
        \begin{subfigure}[b]{0.5\textwidth}
                \centering
                \includegraphics[width=0.94\textwidth]{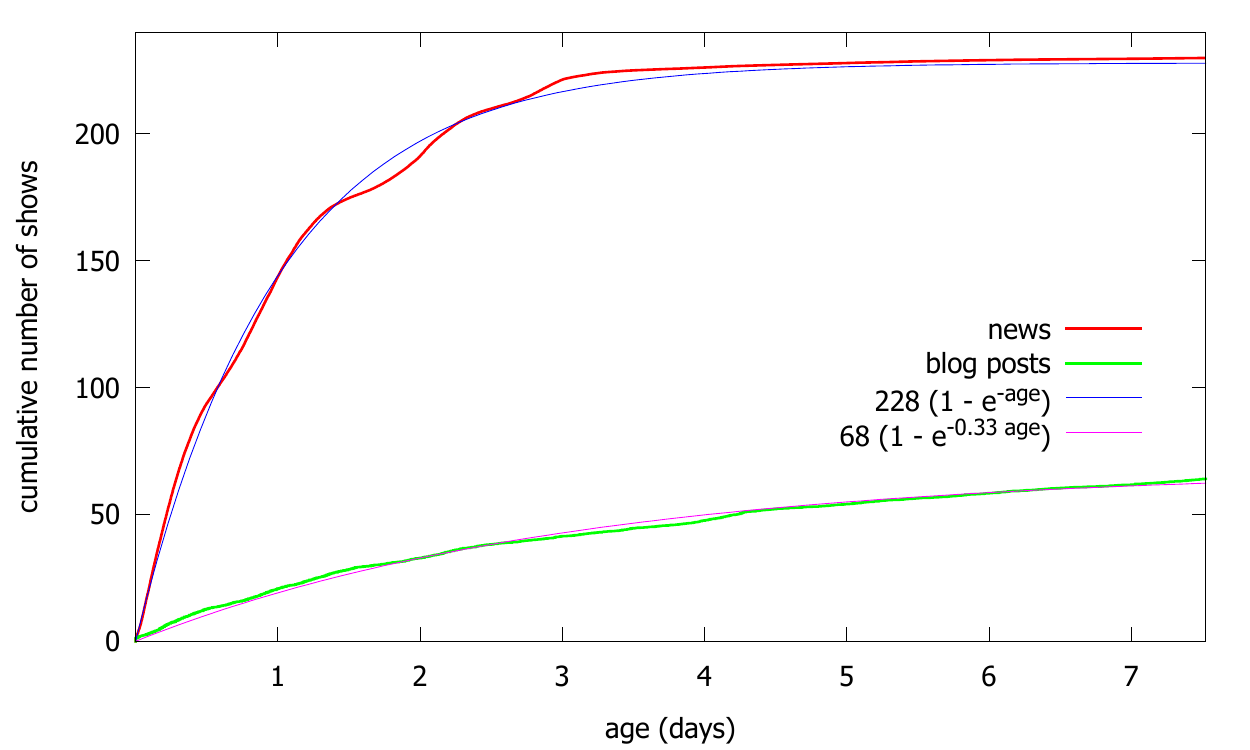}
                \caption{Shows}
                \label{fig:avg_shows}
        \end{subfigure}%
        \begin{subfigure}[b]{0.5\textwidth}
                \centering
                \includegraphics[width=0.94\textwidth]{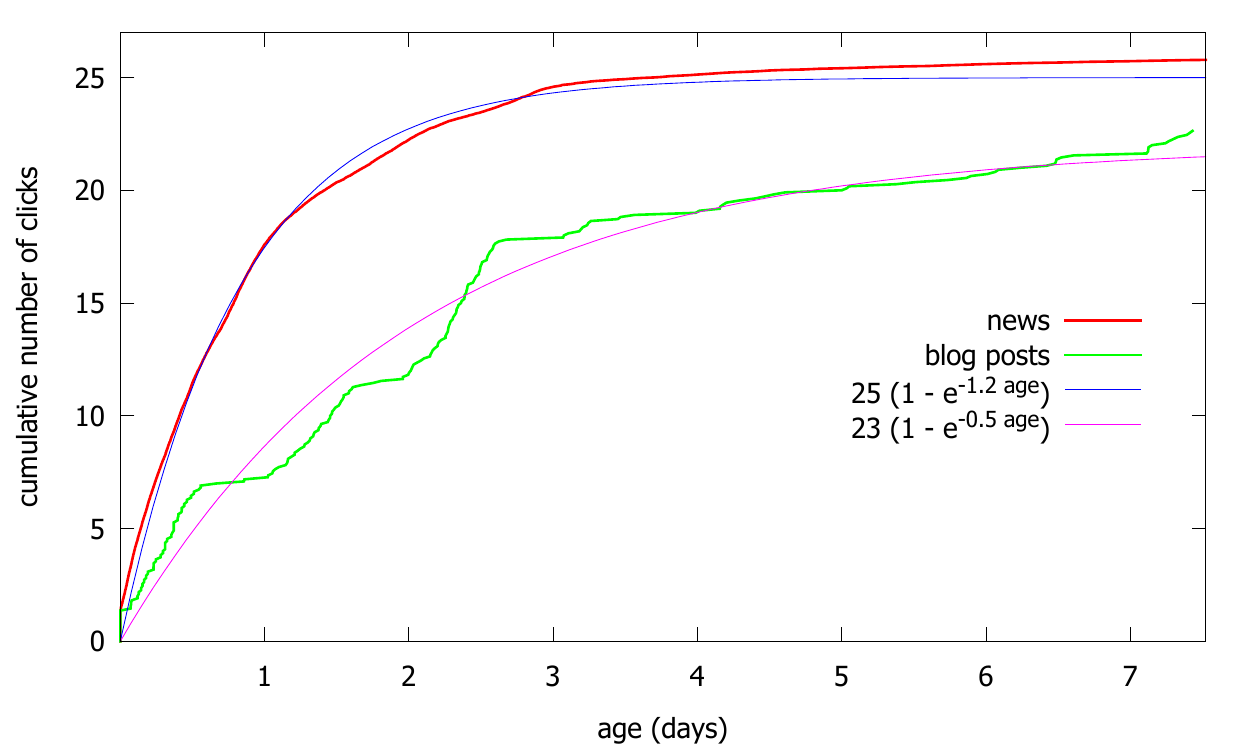}
                \caption{Clicks}
                \label{fig:avg_clicks}
        \end{subfigure}
        \caption{Average cumulative number of shows and clicks depending on page's age}
  \end{figure*}

  However, for our crawling algorithm, we do not want to wait for such
  a long period of time to be able to use the profit of freshly
  crawled pages in order to quickly adapt to changes of content
  sources' properties.  Of course, we do not know the function
  $P_i(\Delta t)$ for pages that just appeared, but we can try to
  predict it.  It is natural to expect that pages of similar nature
  exhibit similar distributions of user interest. In order to
  demonstrate this, on Figure~\ref{fig:avg_shows} and
  Figure~\ref{fig:avg_clicks} we plot respectively the average number
  of cumulative shows and clicks depending on the page's age for all pages
  published on a news site and a blog (both being randomly chosen)
  over $1$ week.
We can see that almost all clicks and shows appear in the first week of
a page's life, and that the dependency of
the cumulative number of clicks (shows) gathered by a page
on the page's age is pretty well
described by the function: $P \left( 1 - e^{- \mu \cdot \Delta
    t} \right)$,
where $P$ is the total number of clicks (shows) a page gathers during its life.
We thus propose the following approximation of the profit $P_i(\Delta t)$ (i.e., the number of future clicks):
$$
P_i(\Delta t) \approx P_i \cdot e^{- \mu_i \cdot \Delta t},
$$
where the \textit{rate of decay} $\mu_i$ and the \textit{profit} $P_i$
are content-source-specific and should be estimated using historical data (see Section~\ref{p&mu}
for details).
We use this approximation in Section \ref{Algo} in order to analyze the problem under consideration theoretically.

\section{Content sources}\label{ContentSources}\label{HowToFind}

In this section, we show that most ephemeral new content can indeed be
found at a small set of content sources and then describe a simple
procedure for finding such a set, that fits our use case.

\subsection{Analysis of content sources}

Our hypothesis is that one can find the most of ephemeral new pages appearing
on the Web at a small set of content sources, but links from these
sources to new pages are short living so a crawler needs to frequently
recrawl these sources to avoid missing links to new pages, especially to
high-quality pages.

In order to validate this hypothesis about content sources, we need to
follow the evolution over time of the link structure of the Web, to
understand which content sources refer which new pages as they appear.
Our Web crawler logs could be used for this, but there are two main
issues with this approach: 1) keeping the full history of new pages
linked from each content source, even for some small time period, is
impractical due to resource constraints; and 2) existing crawlers do not
revisit each content source often enough to provide more than a really
partial view of the evolution of the link structure of the Web.

\begin{figure}
\begin{center}
\includegraphics[width = 0.48\textwidth]{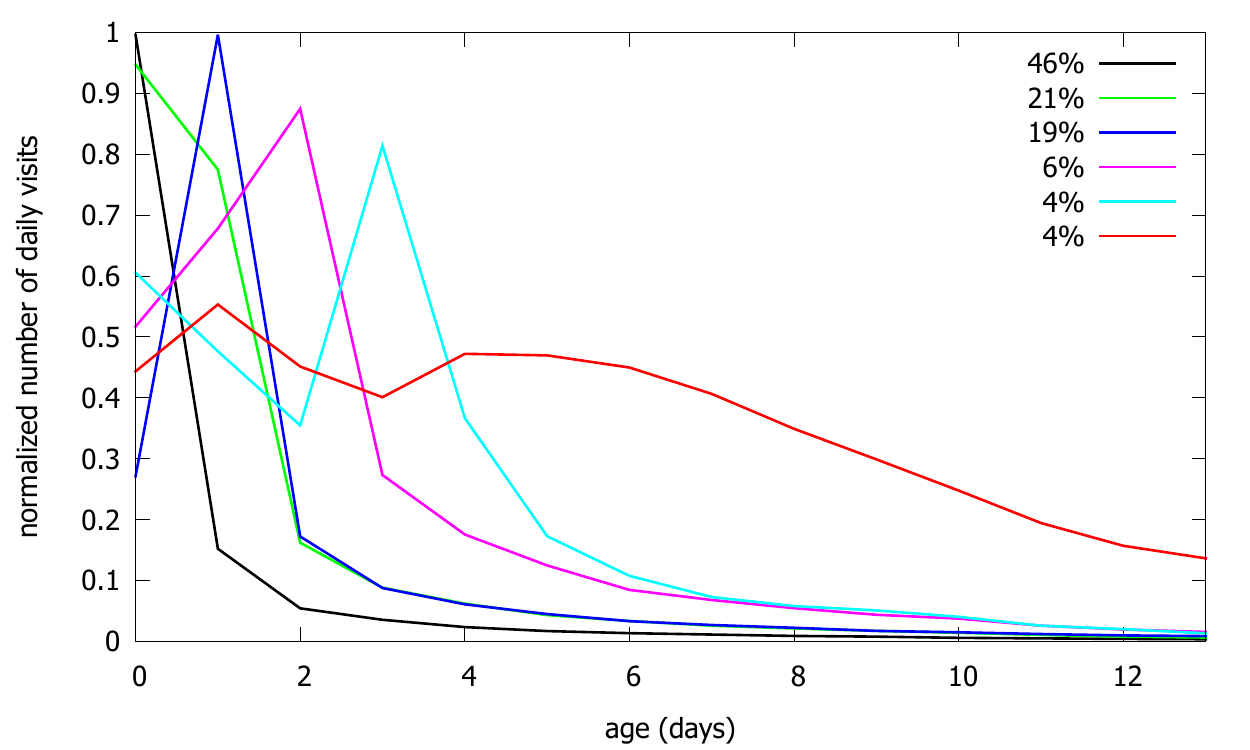}
\end{center}
\caption{User interest patterns for new pages.}
\label{fig:centroids-6}
\end{figure}

So, instead, we used the toolbar logs continuously collected at Yandex (Russia's most popular search engine) to monitor user visits to
Web pages.
In this way, we can easily track the appearance of new pages that are of interest to at least one user,
know which content sources referred them, and also follow the
evolution of user interest over time for each of these new pages. This
data is a representative sample of the Web as this toolbar is used by
millions of people across different countries. But we cannot use this
data in the algorithm itself since it is not available in all
countries, and thus we use it only in order to validate our hypothesis
about the existence of relatively small set of content sources.

Using this toolbar data, we randomly sampled 50K pages from the set of
new pages that appeared over a period of one week and were
visited by at least one user. These pages were
distributed over $\thicksim1.6$K different hosts.  For each page, we
computed its number of daily visits for a period of two weeks after it
appeared. Then, using this 14-dimensional (one per day) feature vector
(scaled to have its maximum value equal to 1), we clustered these pages
into 6 clusters by applying the k-means method\footnote{\url{http://en.wikipedia.org/wiki/K-means_clustering}}.
Let us note that when we tried less clusters, non-ephemeral pages were not assigned to one cluster.
Finally, we obtained only $\thicksim4$\% non-ephemeral pages.
The percentage of new pages that are ephemeral (and were visited at least once) for this week is thus 96\%, which is really significant.
Centroids of these clusters are plotted on Figure~\ref{fig:centroids-6}
(in Section~\ref{Intro} we showed only two of them).

Our toolbar logs also contain, in most cases, the referrer page for
each recorded visit to a page, i.e., the source page from which the user came
to visit this target page.  We extracted all
these \textit{links} (users transitions between pages) found in the logs pointing to one of the ephemeral new
pages in our sample over the same period of one week plus several days (for the most
recent pages to become obsolete), and obtained ${\thicksim750}$K
links.

\begin{figure}
\includegraphics[width=0.46\textwidth]{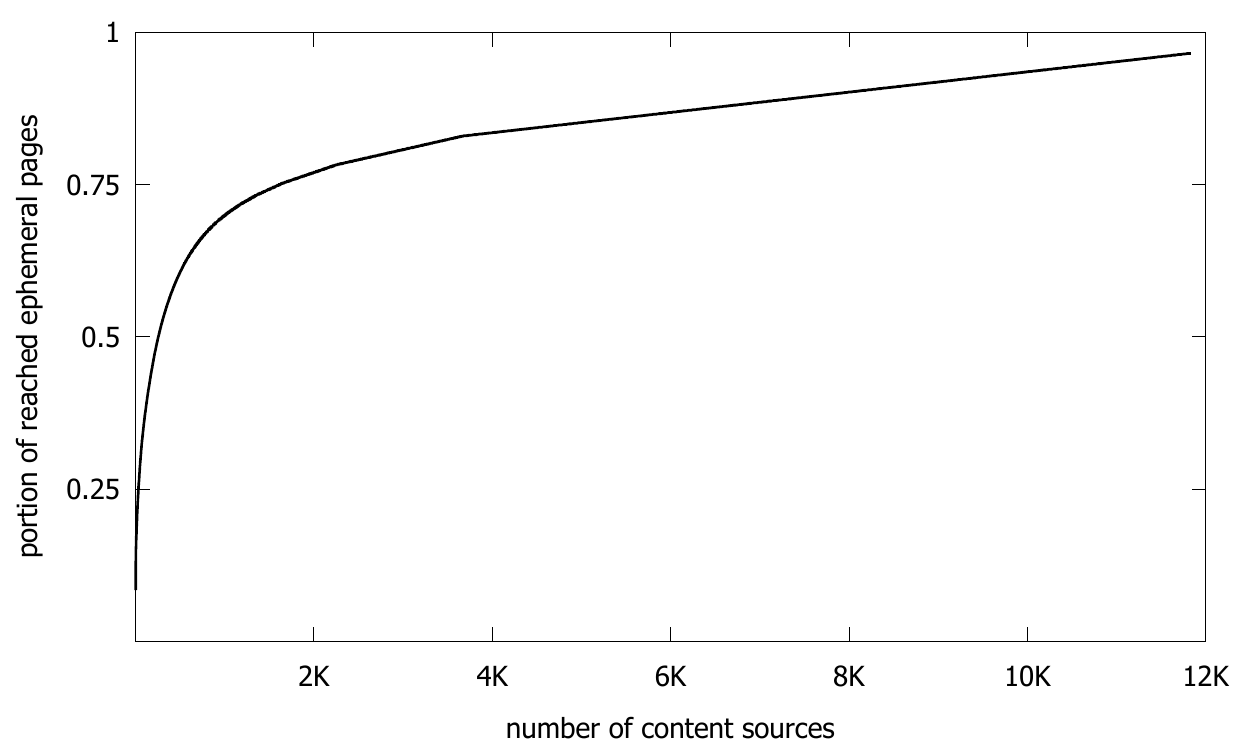}
 \caption{Share of ephemeral new pages reached depending on the average number of content sources per host.}
\label{fig:hubs}
\end{figure}

Using these links, we studied the percentage of ephemeral new pages
reached depending on the number of content sources. We want to find
the smallest set of content sources that allows to reach the most of new pages
and thus proceed as follows (in a greedy manner).  We first take the
source page, which allows to reach the most of pages, then
remove it and all covered pages, then select the second one in the
same way, and so on.  We see on Figure~\ref{fig:hubs}, that only 3K
content sources are required to cover 80\% of new content, which 
validates our hypothesis about content sources. Interestingly, 42\% of
these 3K content sources are main pages of web sites, while 44\% are
category pages, which can be accessed from the main page. So, overall,
86\% of them are at most 1 hop away from the main page. 


%

\subsection{Procedure to find content sources}

Now, we need to understand how to effectively find these content
sources at scale without relying on toolbar logs, which, as said,
are not available world-wide.
Our crawling algorithm (described in the next section) focuses on
high-quality ephemeral new content. Therefore, even if content sources
that produce low quality content or almost no new content at all are
given to this algorithm, they will almost never be crawled or crawled
just in case much later when some spare resources are available (see
Section \ref{Algo}). We are thus, to some extent, only interested in
recall when finding such sources here, i.e., to get most of them,
which makes this task much easier to solve in practice.



Analyzing the set of content sources discovered using toolbar data, we
noticed that 86\% of content sources that we found are actually at most 1 hop
away from the main page of their host, as said. The following procedure
will thus yield a relatively small set of content sources that
generate most of the ephemeral new content on a given set of hosts.

\begin{procedure}
\SetKwInOut{Input}{input}\SetKwInOut{Output}{output}
\Input{A list of hosts that generate new content}
\Output{A list of content sources}
\BlankLine
\begin{enumerate}
\item Crawl the main page of each host (once) and keep pages linked from it (all
  pages 1-hop away from the main page);
\item Select the main page, and all found pages older than few
  days, by using historical data, as content sources (as new pages
  are almost never content sources).
 \end{enumerate}
\caption{to find content sources with good recall()}\label{algo_find_cs}
\end{procedure}


Let us note that this procedure is easy to run periodically to refresh
the set of content sources, and to find new content sources.
It is possible to use URL patterns as described
in \cite{Liu} in order to get a better precision, but this
optimization is not required here, because our crawling
algorithm optimizes precision itself by avoiding to crawl
low-quality content sources (see Section \ref{Algo}).

The input list of hosts for this procedure can be obtained
during a standard web crawling routine
by ranking found hosts by their tendency to generate new content.
This simple method also fits our usage scenario considering that, as
said, only recall is important for us when finding content sources as input for our algorithm.


\section{Optimal crawling \\ of content sources}\label{Algo}
In this section, we assume that we are given a relatively small set of
content sources, which regularly generate new content (see Section
\ref{HowToFind} for the procedure to select such a set). Our current aims are to
(1) find an optimal schedule to recrawl content sources in order to
quickly discover high-quality ephemeral new pages,
and (2) understand how to spread resources between crawling new pages
and recrawling content sources.

First, we analyze this problem theoretically and find an optimal
solution. Then, we describe an algorithm, which is based on this
solution.

\subsection{Theoretical analysis}\label{Theory}



Assume that we are given a set of content sources $S_1, \ldots, S_n$.
Note that the rate of new content appearance may differ from source to source.
For example, usually there are much more news about politics than about art,
and therefore, different categories of a news site generate new content with different rates.
Let $\lambda_i$ be \textit{the rate of new links appearance} on the source
$S_i$, i.e., the average number of links to new pages, which appear in
one second.

Let us consider an algorithm, which recrawls each source $S_i$
every $I_i$ seconds, discovers links to new pages, and also crawls all
new pages found. We want to find a schedule for recrawling
content sources, which maximizes the overall quality $Q$ (see Equation (\ref{Q})), i.e., our aim
is to find optimal values of $I_i$. Suppose that our infrastructure allows
us to crawl $N$ pages per second ($N$ can be non-integer).
Due to these resource constraints, we have the following restriction:
$$
\sum_i \frac{1+\lambda_i I_i}{I_i} \le N.
$$
On average, the number of new links linked from a source $S_i$ is equal to $\lambda_i I_i$
, therefore every $I_i$ seconds we have to crawl $1+\lambda_i I_i$ pages (the source itself and all new pages found).
Obviously, the optimal solution requires to spend all the resources:
\begin{equation}\label{Restriction}
\sum_i \frac{1}{I_i} = N - \sum_i \lambda_i.
\end{equation}
And we want to maximize the overall quality (see Equation~(\ref{Q})), i.e.,
$$
 Q = \sum_i \frac{1}{I_i}  \sum_{j: p_j \in S_i \wedge t_j \in [0, I_i]} P_j(\vartriangle t_j) \rightarrow \max.
$$
Note that this expression is exactly the average profit per second.

\begin{figure}
\begin{center}
\includegraphics[width = 0.46\textwidth]{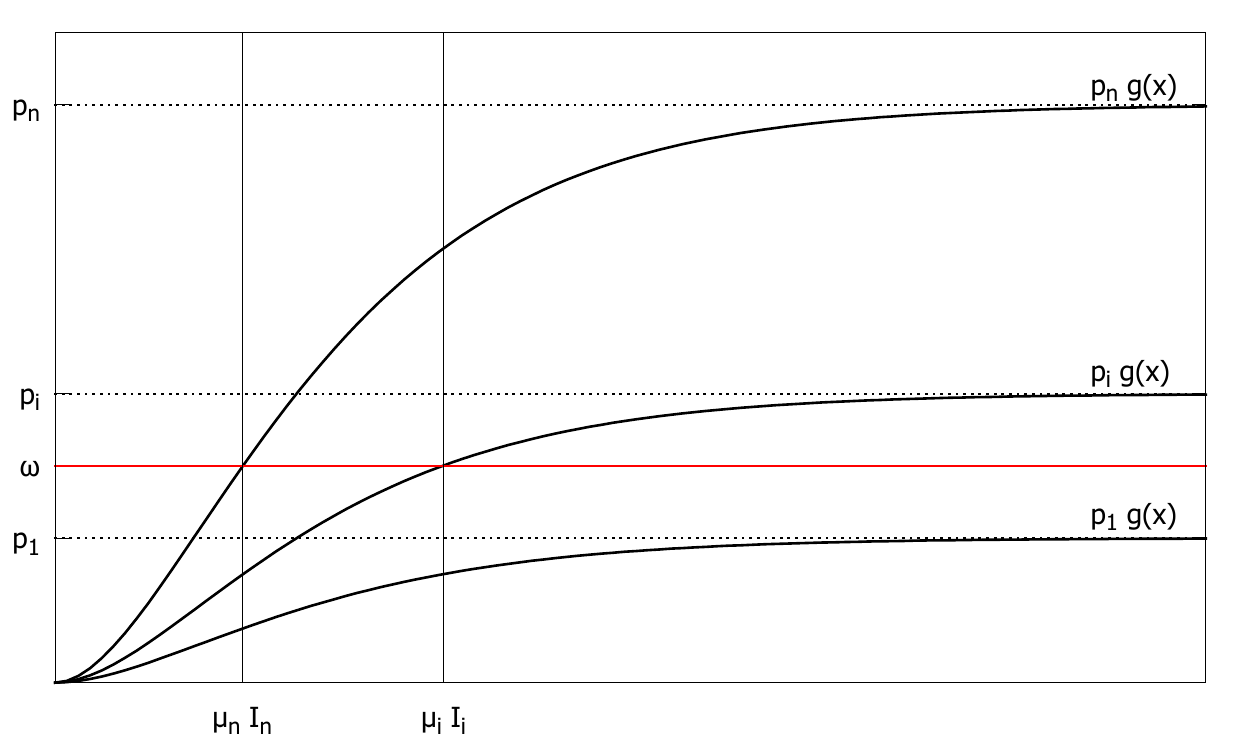}
\end{center}
\caption{Optimizing of $I_i$}
\label{fig:omega}
\end{figure}

Content sources may not be equal in quality, i.e., some
content sources may provide users with better content than others.
We now assume that, on average, the pages from one content source exhibit the same behavior of the profit decay function
and hence substitute $P_j(\Delta t_j)$ by the approximation $P_i e^{-\mu_i \Delta t_j}$ discussed in Section~\ref{Formalization}.
We treat  the profit $P_i$ and the rate of profit decay $\mu_i$ as the parameters of each content source $S_i$.
Thus, we obtain:
$$
Q = \sum_i \frac{P_i}{I_i} \sum_{j=0}^{\lambda_i I_i - 1} e^{-\mu_i \frac{j}{\lambda_i}} =
$$
$$
= \sum_i \frac{P_i}{I_i} \frac{1 - e^{-\mu_i I_i}}{1 - e^{-\frac{\mu_i}{\lambda_i}}} =
\sum_i p_i x_i \left(1 - e^{-\mu_i / x_i}\right),
$$
here $p_i = \frac{P_i}{1 - e^{-\frac{\mu_i}{\lambda_i}}}$ and $x_i = \frac{1}{I_i}$. Without loss of generality, we can assume that $p_1\leq \ldots \leq p_n$.
We now want to maximize $Q(x_1, \dots, x_n)$ subject to (\ref{Restriction}). We use the method of Lagrange multipliers:
\begin{align*}
  \begin{cases}
     p_i \left(1 - e^{-\mu_i / x_i}\right) - \frac{\mu_i p_i}{x_i} e^{-\mu_i / x_i} = \omega & i = 1, \dots, n  \, ,\\
  \sum_i x_i = N - \sum_i \lambda_i \, ,        &
  \end{cases}
\end{align*}
where $w$ is a Lagrange multiplier.

Note that $I_i = \frac{1}{x_i}$, so we get:
\begin{align}\label{Solution2}
  \begin{cases}
     p_i \left( 1 - (1 + \mu_i I_i) e^{-\mu_i I_i} \right) = \omega,        & i = 1, \dots, n \, , \\
  \sum_i \frac{1}{I_i} = N - \sum_i \lambda_i \, .        &
  \end{cases}
\end{align}
The function $g(x) = \left( 1 - (1 + x) e^{-x} \right)$ increases
monotonically for ${x>0}$ with ${g(0) = 0}$ and ${g(+\infty) = 1}$. If we are
given ${0 < \omega < p_i}$, then we can find the unique value $\mu_i
I_i = g^{-1}( \frac{\omega}{p_i} )$ as shown in Figure~\ref{fig:omega}. One can easily compute $g^{-1}$ using a binary
search algorithm\footnote{\url{http://en.wikipedia.org/wiki/Binary_search_algorithm}}.
Note that bigger values of $\omega$ lead to bigger values of
$\mu_i I_i$. That is why $\sum_i \frac{1}{I_i}$ is a monotonic
function of $\omega$ and we can, here also, apply a binary search algorithm (see Algorithm \ref{algo_disjdecomp}) to
achieve the condition $\sum_i \frac{1}{I_i} = N - \sum_i \lambda_i$.

\IncMargin{1em}
\begin{algorithm}
\SetKwInOut{Input}{input}\SetKwInOut{Output}{output}
\Input{profits $P_i$, rates of profit decay $\mu_i$, intensities of new links appearance $\lambda_i$, number of crawls per second $N$, precision $\varepsilon$}
\Output{optimal recrawl intervals $I_i$}
\BlankLine
$\omega_l \longleftarrow  0; \omega_u \longleftarrow  p_n = \frac{P_n}{1 - e^{-\frac{\mu_n}{\lambda_n}}}$\;
\While{ $\left|\sum_{i:I_i \neq \infty} \frac{1}{I_i} - N + \sum_{i:I_i \neq \infty} \lambda_i \right| > \varepsilon$}{\
  $\omega \longleftarrow  \frac{ \omega_u + \omega_l } {2}$\;
  $I_i \longleftarrow \frac{1}{\mu_i} g^{-1}\left( \frac{\omega}{p_i} \right)
   = \frac{1}{\mu_i} g^{-1}\left( \omega \frac{1 - e^{-\frac{\mu_i}{\lambda_i}}}{P_i} \right)$\;
  \BlankLine
  \If{ $\sum_{i:I_i \neq \infty} \frac{1}{I_i} < N - \sum_{i:I_i \neq \infty} \lambda_i$} {
     $\omega_u \longleftarrow \omega$\;
  } \Else {
     $\omega_l \longleftarrow \omega$\;
  }
}
\caption{Find an optimal recrawl schedule}\label{algo_disjdecomp}
\end{algorithm}\DecMargin{1em}

Let $\omega_l$ and $\omega_u$ be, respectively, the lower and upper bounds for $\omega$. At the first step, we can put $\omega_l=0$ and $\omega_u=p_n$. Indeed, $p_n$ is the obvious upper bound for $\omega$, since in this case we do not crawl any content source. At each step of the algorithm, we consider $\omega = \frac{ \omega_u + \omega_l } {2}$. For this value of $\omega$, we recompute intervals $I_i=\frac{1}{\mu_i} g^{-1}\left(\frac{\omega}{p_i} \right)$. Note that if we get $\omega > p_j$ for some $j$, then $I_j = \infty$ and we never recrawl this content source. After that, if $\sum_{i:I_i \neq \infty} \frac{1}{I_i}<N - \sum_{i:I_i \neq \infty} \lambda_i$,
then we can put $\omega_u=\omega$, since it is an upper bound. If not, we put $\omega_l=\omega$. We proceed in this way until we reach the required precision $\varepsilon$.

The value of $\omega$ may be interpreted as the threshold we apply to content sources' utility. Actually, we can find the minimal crawl rate required for the optimal crawling policy not to completely refuse to crawl content sources with the least utility.


We completely solved the optimization problem for the metric suggested in Section \ref{Formalization},
the solution of (\ref{Solution2}) is theoretically optimal (we use the name \textit{ECHO-based crawler} for the obtained algorithm, where ECHO is an abbreviation for Ephemeral Content Holistic Ordering).
However, some further efforts are required in order to make this algorithm practically useful. There are parameters, which we need to estimate for each source: the profit $P_i$, the rate of profit decay~$\mu_i$, and the rate of new links appearance~$\lambda_i$.
In the following section, we describe a practical crawling algorithm, which is based on our theoretical results.


\subsection{Implementation}\label{Realization}

Let us describe a concrete algorithm based on the results of Section
\ref{Theory}.  First, we use the results from Section \ref{HowToFind}
to obtain an input set of content sources.  Then, in order to apply
Algorithm \ref{algo_disjdecomp} for finding an optimal recrawl
schedule for these content sources, we need to know for each source
its profit $P_i$, the rate of profit decay~$\mu_i$, and the rate of
new links appearance~$\lambda_i$.  We propose to estimate all these
values dynamically using the crawling history and search engine logs.
Since these parameters are constantly changing, we need to periodically
re-estimate time intervals $I_i$ (see Algorithm
\ref{algo_disjdecomp}), i.e., to update the crawling schedule. Obviously, the more
often we re-estimate $I_i$, the better results we will
obtain, and the choice of this period depends on the computational
resources available.


Thus, we first discuss how to estimate these sources' characteristics (Sections \ref{p&mu} and \ref{lambda}) and then how to deal with deviations of content sources'
behavior from our idealistic assumptions to make a practical scheduling algorithm (Section \ref{scheduling}).

\subsubsection{Estimation of profits $P_i$ and \\ rates of profit decay $\mu_i$} \label{p&mu}

For this part, we need search engine logs to analyze the history of clicks on new pages.
We want to approximate the average cumulative number of clicks depending on the page's age by an exponential function.
This approximation for two chosen content sources is shown on Figure~\ref{fig:avg_clicks}.

Let us consider a cumulative histogram of all clicks for all new pages
linked from a content source, with the histogram bin size equals to $D$ minutes.
Let $s_i$ be the number of times all $N$ new pages linked from this content source
were clicked during the first $i D$ minutes after they appeared.
So, $s_i/N$ is the average number of times a new page was clicked during the first $i D$  minutes.

We can now use the least squares method, i.e., we need to find:
\begin{equation}\label{LeastSquares}
\arg\min_{\mu, P} F(P,\mu) = \arg\min_{\mu, P} \sum_i \left(P\left(1-e^{-\mu i D }\right) - \frac{s_i}{N} \right)^2.
\end{equation}

In other words, we want to find the values of $\mu$ and $P$,
that minimize the sum of the squares of the differences between
the average cumulative number of clicks and its approximation  $P \left( 1 - e^{-\mu i D} \right)$.
It is hard to find an analytical solution of (\ref{LeastSquares}), but we can use the gradient descent method to solve it:

\IncMargin{1em}
\begin{algorithm}\label{GradientDescent}
\SetKwInOut{Input}{input}\SetKwInOut{Output}{output}
\Input{histogram bin size $D$, cumulative number of clicks $s_i$,
number of new pages found $N$,
precision $\varepsilon$, step size $\gamma$, initial values $P_{init}$ and $\mu_{init}$}
\Output{profit $P$, rate of profit decay $\mu$}
\BlankLine
$P_{old} \longleftarrow  0; P \longleftarrow  P_{init}; \mu_{old} \longleftarrow  0; \mu \longleftarrow  \mu_{init}$;
\While{$\max\{|P_{old}-P|,|\mu_{old}-\mu|\} > \varepsilon$}{
$P_{old} \longleftarrow P;$\\
$\mu_{old} \longleftarrow \mu;$\\
$P \longleftarrow P_{old} - \gamma  {\partial F \over \partial P}(P_{old},\mu_{old});$\\
$\mu \longleftarrow \mu_{old} - \gamma {\partial F \over \partial \mu}(P_{old},\mu_{old});$\footnotemark[4]
}
\caption{Estimate profit decay function}
\end{algorithm}\DecMargin{1em}
\footnotetext[4]{${\partial F \over \partial P}(P,\mu) =
2 \sum_i \left( P \left( 1 - e^{-\mu iD} \right) - \frac{s_i}{N}\right)\left( 1 - e^{-\mu iD} \right)\\
{\partial F \over \partial \mu}(P,\mu) =
2 \sum_i \left( P \left( 1 - e^{-\mu iD} \right) - \frac{s_i}{N}\right) iPD e^{-\mu iD}$}

From the production point of view, it is very important to decide, how
often to push data from search engine logs to re-estimate the values of
$\mu_i$ and $P_i$ as it is quite an expensive operation.
We denote this \textit{logs push period} by $L$. In
Section \ref{Exp}, we analyze how the choice of $L$ affects the
performance of the algorithm.

\subsubsection{Estimation of the rate of new links \\appearance $\lambda_i(t)$}\label{lambda}

The rate of new links appearance $\lambda_i(t)$ may change during the day
or during the week. We thus dynamically estimate this rate for each
content source. In order to do this, we use historical data: we
consider the number of new links found at each content source during
the last $T$ crawls.
We analyze how different values for $T$ affect the performance of the algorithm in Section~\ref{Exp}.

\subsubsection{Scheduling}\label{scheduling}

Finally, in order to apply our algorithm,
we should solve the following problem: in reality the number of new links that appear on a content source during a fixed time period is random
and we cannot guarantee that we find exactly $\lambda_i I_i$ new links after each crawl.
We can find more links than expected after some recrawl and if we crawl all of them,
then we will deviate from the schedule. Therefore, we cannot both stick to the schedule
for the content sources and crawl all new pages.
So we propose the two following variants to deal with these new pages,
that we cannot crawl without deviating from the schedule.

\textbf{ECHO-newpages.} 
In order to avoid missing clicks,
we always crawl newly discovered pages right after finding them.
If there are no any new pages in the crawl frontier,
we try to come back to the schedule.
We crawl the content source, which is most behind the schedule,
i.e., with the highest value of $I_i' / I_i $, where $I_i'$
is time passed after the last crawl of the $i$-th content source.



\textbf{ECHO-schedule.}
We always crawl content sources with
intervals $I_i$ and when we have some resources to crawl new
pages, we crawl them (most recently discovered first).


We compare these two variants experimentally in the next section.

\subsubsection{Possible production architecture}

\begin{figure}
\begin{center}
\includegraphics[width = 0.46\textwidth]{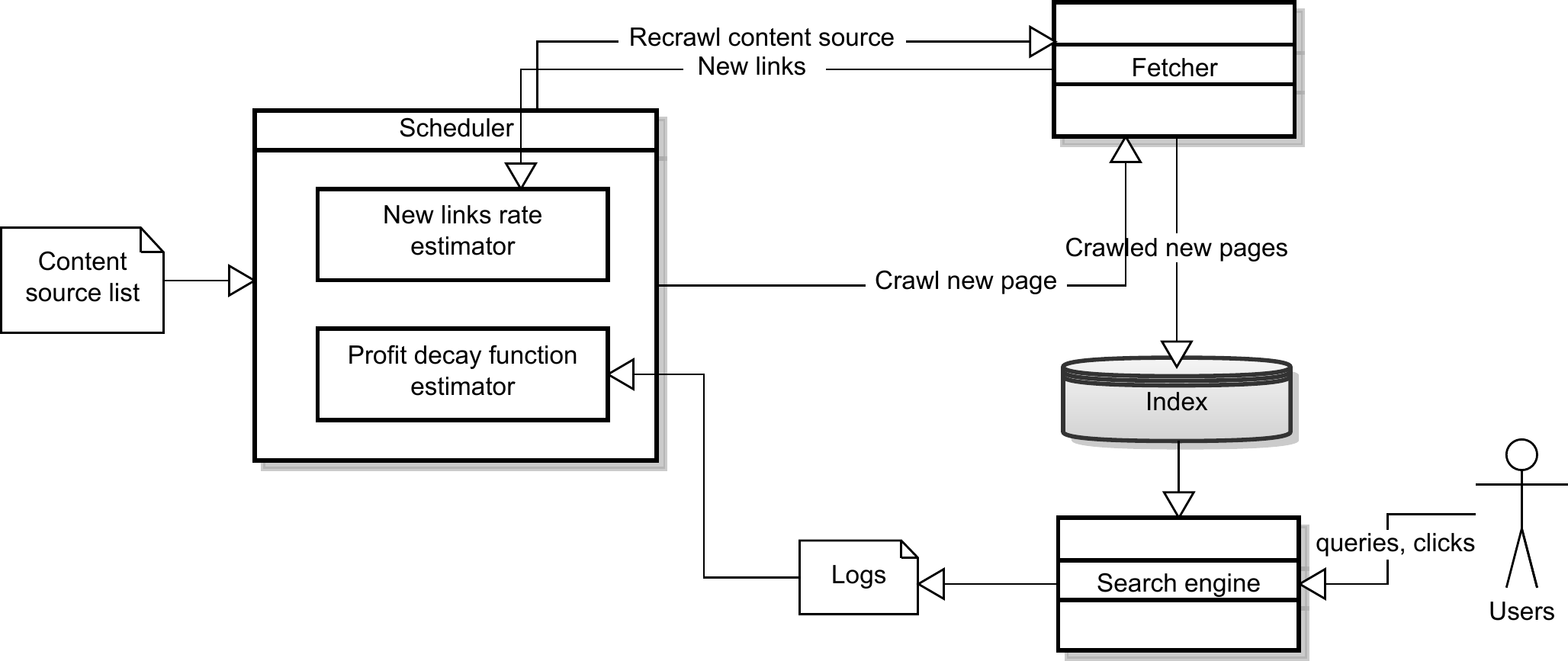}
\end{center}
\caption{Possible production architecture.}
\label{fig:architecture}
\end{figure}

We finish this section by presenting a possible production
architecture for our algorithm (see Figure~\ref{fig:architecture}) to
emphasize that it is highly practical to implement it in a
production system. Initially, a set of content sources with good recall of
new pages linked from these sources is created using the procedure
described in Section~\ref{ContentSources}. This procedure must be run
periodically to refresh this set and include new content
sources. Then, \textit{Scheduler} finds the optimal crawling schedule
for content sources, while \textit{Fetcher} crawls these sources
according to this schedule. \textit{Scheduler} dynamically estimates
the rate of new links appearance for content sources and also
estimates the profit decay function using the number of clicks from
the search engine logs.  Given a fixed ranking method, the number of
clicks measures the direct effect of the crawling algorithm on the
search engine's performance. \textit{Scheduler} dynamically uses this
feedback in order to improve the crawling policy.


\section{Experiments}\label{Exp}

In this section, we compare our algorithm with some other crawling
algorithms on real-world data.

\subsection{Data}\label{Data}

Since it is impractical and unnecessary to conduct research experiments at a production scale,
we selected some sites that provide a representative sample of
the Web, on which we performed our experiments.


We selected the top 100 most visited Russian news sites and the top 50 most
visited Russian blogs  using publicly available data from trusted sources\footnote[5]{\url{http://liveinternet.ru/rating/ru/media/}  \\ \url{http://blogs.yandex.ru/top/}}.
We consider these web-sites to be a representative sample of the Web
for our task as they produce 5-6\% out of the $\thicksim500$K new pages (visited by at least one user) that appear in this country daily
(we estimated this second value using toolbar logs).
For each such site, we applied the procedure described in Section \ref{ContentSources}
and obtained about 3K content sources.

Then, we crawled each of these content sources
every 10 minutes for a period of 3 weeks (which is frequent enough to be able to collect all
new content appearing on them before it disappears). The discovery
time of new pages we observed is thus at most delayed by these 10
minutes.  We considered all pages found at the first crawl of each
source (each content source was crawled $\thicksim3K$ times) to be old and discovered $\thicksim415$K
new pages during these $3$ weeks.  Keeping track of when links to new
pages were added and deleted from the content sources, we created a
dynamic graph that we use in the following experiments. This graph
contains $\thicksim2.4$M unique links.

Additionally, we used search engine logs of a major search engine to collect user clicks for each of
the newly discovered pages in our dataset for the same period of $3$
weeks plus $1$ week for the most recent pages to become obsolete. We
observed that $\thicksim20$\% of the pages were clicked at least once during this
4 weeks period.

\subsection{Simplifications of the algorithm}\label{Benchmarks}

We compare the algorithm suggested in Section \ref{Algo} with several
other algorithms.  There are no state-of-the-art algorithms for the
specific task we discuss in this paper, but one can think of several
natural ones:

\begin{itemize}
\item \textbf{Breadth-first search (BFS)} We crawl content sources
  sequentially in some fixed random order. After crawling each source,
  we crawl all new pages linked from this source, which have not been
  crawled yet.


\end{itemize}

We also compare our algorithm with the following simplifications to
understand the importance of 1) the holistic crawl ordering and 2) the
usage of clicks from search engine logs.
\begin{itemize}
\item \textbf{Fixed-quota}
  This algorithm is similar to \textit{ECHO-\\schedule}, but we use a fixed
  quota of $\frac{1}{2}$ for recrawling content sources and for crawling
  new pages that have not been crawled before.
\item \textbf{Frequency} This algorithm is also
  similar to \textit{ECHO-schedule}, but we do not use clicks from search
  engine logs, i.e., all content sources have the same quality and
  content sources are ordered only by their frequency of new pages
  appearance.
\end{itemize}

We also propose a simplification of our algorithm, based
on Section \ref{Algo}, which could be much easier to implement in
a production system.

\begin{itemize}
\item \textbf{ECHO-greedy} We crawl the content source with the
  highest expected profit, i.e., with the highest value of $\lambda_i
  P_i I'_i$, where $I'_i$ is the time passed since the last crawl of
  the content source, $\lambda_i$ is its rate of new links
  appearance, and $P_i$ is the average profit of new
  pages linked from the content source. Then, we crawl all new pages
  linked from this source, which have not been crawled yet, and repeat
  this process. 
\end{itemize}

\subsection{Results}

\subsubsection{Experimental scheme}
In this section, we experimentally investigate the influence of parameters on our algorithm's performance and compare the algorithm with the approaches from Section \ref{Benchmarks} on real-world data.

We simulated, for each algorithm, the crawl of the dynamic graph
described in Section \ref{Data}, using the content sources as seed
pages. Each algorithm can thus, at each step, decide to either crawl a
newly discovered page or to recrawl a content source in order
to find new pages.

\begin{figure}
\begin{center}
\includegraphics[width = 0.46\textwidth]{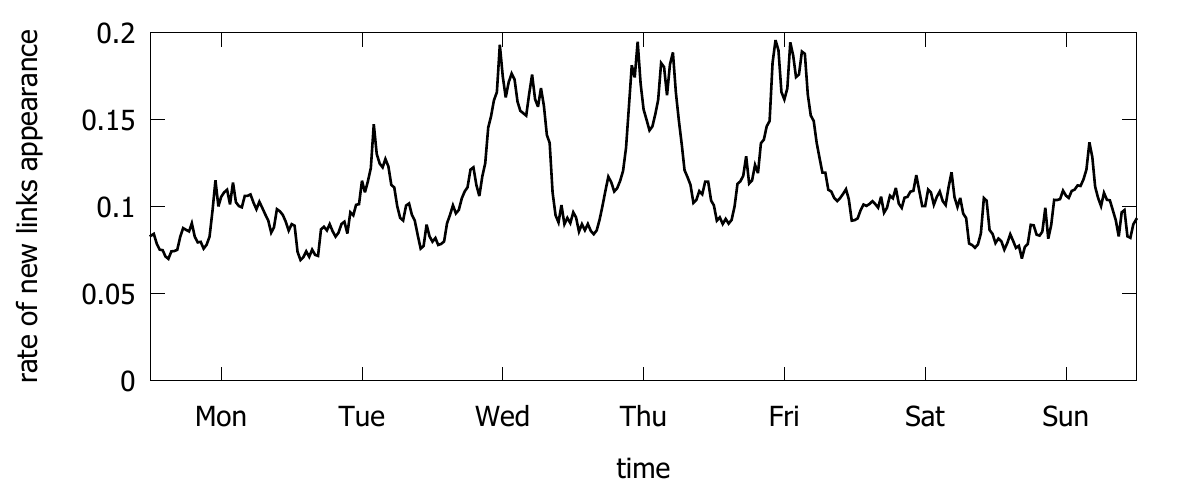}
\end{center}
\caption{Estimation of the rate of new content appearance.}
\label{fig:lambda}
\end{figure}

In the following experiments analyzing parameters influence, we used the
crawl rate per second $N = 0.1$.  This crawl rate is enough to crawl a
significant fraction of the new pages as shown on
Figure~\ref{fig:lambda}, but is not too high to let BFS algorithm crawl all new pages (which is highly unrealistic in a production
context). We then also use two other crawl rates $N = 0.05$ and $N =
0.2$ per second to investigate the influence of this value.

\subsubsection{Influence of parameters}\label{influence}

We apply Algorithm~\ref{algo_disjdecomp}
to re-estimate $I_i$ values every 30 minutes, which is frequent enough
so that smaller intervals have almost no influence on its
performance, and which is also realistic in a production context.  We
also set the bin size $D$ used in Algorithm 2 to 20 minutes,
which is good enough to have robust estimations of $P_i$ and
$\mu_i$, as, typically, the profit decay function $P_i(\triangle t)$
does not change significantly in such a small time period.  We do not
study in details, here, the influence of these two parameters due to space
constraints as, according to the experiments we performed, it is negligible, for values below these realistic choices, in comparison with other parameters.

Besides that, we need default values for profits $P_i$ as we start
crawling without knowing anything
about the quality of each content source.  Please, note that
we need to use \textit{pessimistic} default values because we
want to avoid crawling low quality sources too frequently, while we
do not have enough feedback to have precise estimations.  We cannot
use $P_{default} = 0$ as according to Algorithm~\ref{algo_disjdecomp},
we do not crawl content sources with zero profit, so, we used some
small non-zero value $P_{default}=0.01$.

\begin{figure}
\begin{center}
\includegraphics[width = 0.46\textwidth]{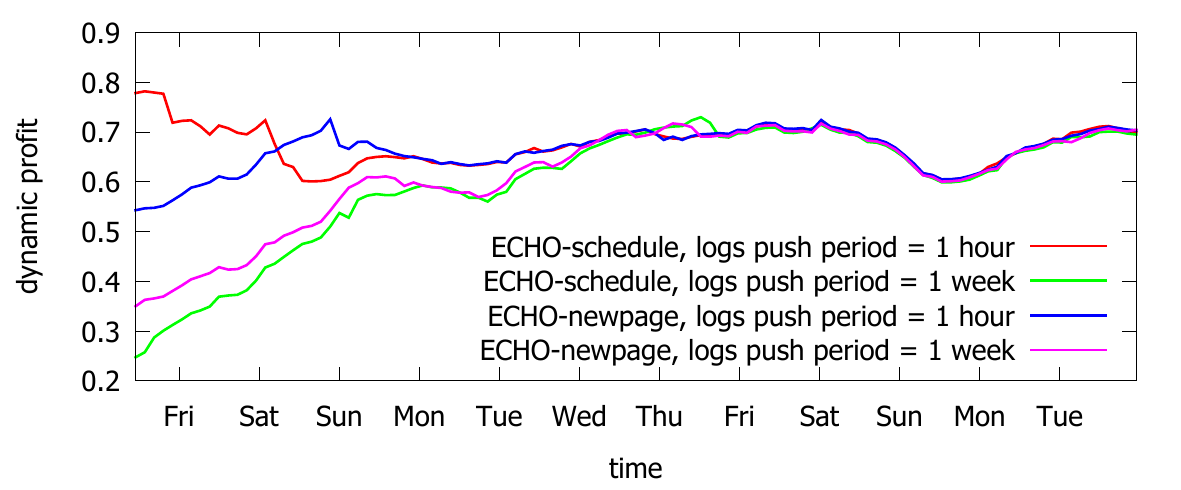}
\end{center}
\caption{Dynamic profit for a 1-week time window.}
\label{fig:profit_7}
\end{figure}

We compared the two variants of ECHO-based crawler from
Section~\ref{scheduling} with different values for: 1) the crawl
history size used to estimate the rate of new links appearance
$\lambda_i(t)$ discussed in Section \ref{lambda} (from 3 to 10
crawls), and 2) the logs push period $L$, which was described in
Section \ref{p&mu} (we considered 1h, 12h, 24h, and 1 week).
Interestingly, for both variants we noticed no difference, and we
therefore conclude that these parameters do not affect the final
quality of the algorithm in our setup.  On the other hand, the logs
push period has a really big influence during the warm-up period and
the smaller the logs push period, the better the results (see Figure~\ref{fig:profit_7}). There is nothing interesting to observe for the
crawl history size.

Let us also note that the optimal schedule of ECHO-based algorithms almost does not recrawl 70\% of content sources, which means that it does not spend much resources on low quality content sources.



\subsubsection{Comparison with other algorithms}\label{compa}

\begin{table}
\begin{center}
\caption{Average dynamic profit for a 1-week window.}
\label{Table1}
\begin{tabular}{|l|c|c|c|c|}
\hline
   Algorithm & N = 0.05 & N = 0.10 & N = 0.20 \\
\hline
Frequency
& $0.014 \pm 0.004$
 & $0.39 \pm 0.04$
 & $0.61 \pm 0.06$ \\
\hline
BFS
& 0.24$\pm$0.04& 0.46$\pm$0.03  &  0.62$\pm$0.03 \\
\hline
Fixed-quota
& 0.43$\pm$0.04 & 0.59$\pm$0.03  &  0.69$\pm$0.03 \\
\hline
ECHO-greedy
& 0.60$\pm$0.03& 0.68$\pm$0.03  &  0.69$\pm$0.03 \\
\hline
ECHO-schedule
& 0.52$\pm$0.02& \textbf{0.69}$\pm$0.03& \textbf{0.71}$\pm$0.03\\
\hline
ECHO-newpages
& \textbf{0.62}$\pm$0.04&\textbf{0.69}$\pm$0.03& \textbf{0.71}$\pm$0.03\\
\hline
\hline
Upper bound
& \textit{0.72}& \textit{0.72} & \textit{0.72} \\
\hline
\end{tabular}
\end{center}
\end{table}

\begin{figure*}
\begin{center}
\includegraphics[width = 18cm]{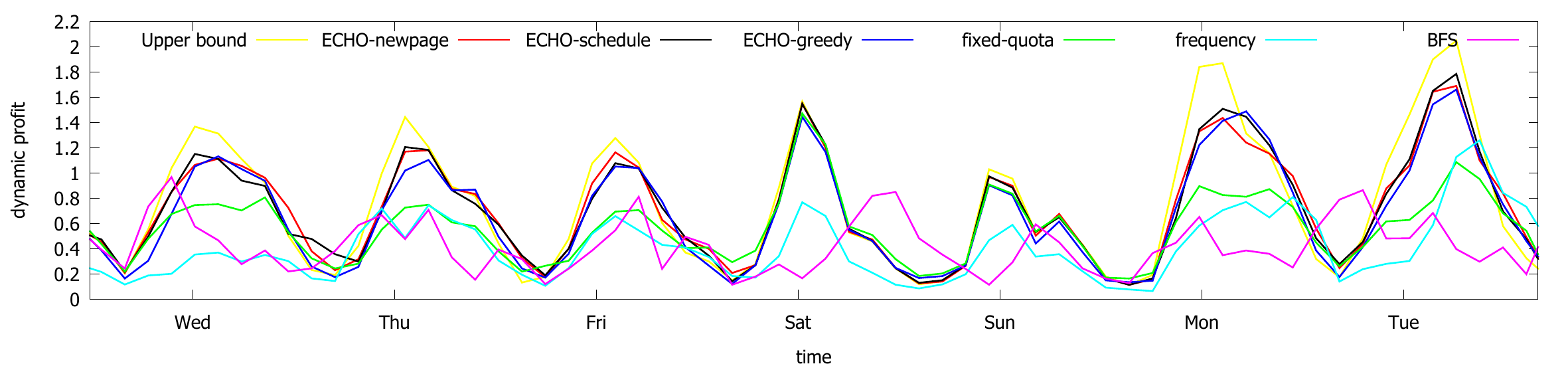}
\end{center}
\caption{Dynamic profit for a 5-hour time window.}
\label{fig:dynamic}
\end{figure*}

Then, we took the crawl history of size 7 and the logs push period of
1 hour (randomly, following the discussion in Section \ref{influence}), and
compared ECHO-based crawlers with other algorithms on three different
crawl rates.  In order to compare our algorithms during the last week  of our
observations (after the warm-up period) we measured the dynamic profit every two minutes using a time window of one week
(enough to compensate daily trends).
Table~\ref{Table1} shows average values and their standard deviations. Note that we also include the upper bound
of algorithms' performance that we computed using BFS algorithm with
an unbounded amount of resources,
which allows to crawl all new pages right after they appear.  This
upper bound therefore does not depend on the crawl rate and equals
$0.72$ of profit per second.


ECHO-newpages shows the best results, which are really close to the upper
bound, although the crawl rate used is much smaller than the rate of
new links appearance.  This means that our algorithm effectively
spends its resources and crawls highest quality pages first.
Note that
the smallest crawl rate that allows BFS to reach 99\% of the upper
bound is 1 per second (this value is measured, but not present in the table), as BFS wastes lots of resources recrawling
content sources to find new pages, while ECHO-newpage and ECHO-schedule reach this bound with crawl rate
0.2 per second (see the last column of the table).


Note that the profit of ECHO-greedy is also high.  This fact can be a
good motivation for using it in a production system, where ease of
implementation is a strong requirement (as it is much easier to
implement).  First, it only requires a priority queue of content
sources rather than a recrawl schedule updated using the binary search
method from Algorithm~\ref{algo_disjdecomp}.  Second, it does not use
$\mu_i$, so $P_i$ is thus simply the average number of clicks on pages
linked from the $i$-th content source, and can therefore be computed
easier than by using the gradient descent method from Algorithm~2.

Let us show a representative example (at $N = 0.1$) demonstrating the advantage of
ECHO-based algorithms over the baselines (see
Figure~\ref{fig:dynamic}). One can observe that ECHO-based algorithms
perform the best most of the time. It is interesting to note though
that during the night BFS shows better results.  It happens as BFS is
``catching up'' by crawling pages, which were crawled by other
algorithms earlier.  This follows how the dynamic profit is defined:
we take into account the profit of the pages, which were crawled
during the last 5 hours. We also see that the algorithm with fixed quota for crawl and recrawl
perform well during the weekend because less new pages appear
during this period and the crawl rate we use is thus enough to crawl
practically all good content without additional optimizations.



\section{Related work}\label{RelatedWork}

Most papers on crawling are devoted to discovering new pages or
refreshing already discovered pages.  Both of these directions are, to
some extent, related to the problem we deal with, though cannot serve as solutions to it.

\paragraph{Refresh policies}

The purpose of refresh policies is to recrawl known pages that have
changed in order to keep a search engine's index fresh. Usually, such
policies are based on some model, which predicts changes on Web pages.
In pioneering works \cite{Cybenko1,Cybenko2,Cho1,Cho2,Cho3}, the analysis of
pages' changes was made in the assumption of a time-homogeneous
Poisson process, i.e., it was assumed that the pages change rate does
not depend on time. However, in \cite{Cybenko1}, it was noted that
there are daily and weekly trends in the pages change rate. Then, a
history-based estimator, which takes such trends into account, was
proposed in \cite{Matloff}.  A more sophisticated approach based on
machine learning is used in \cite{ContentChange}, where the page's
content, the degree of observed changes and other features are taken
into account.

For our specific task, refresh policies can be used to find links to
ephemeral new pages, that appeared on already known pages (content
sources). So, pages changes are relevant for us only if new links to
such new pages can be found.  Interestingly, this simplifies the
estimation of pages change rate as one can easily understand, given
two successive snapshots of a page, that two new links appeared, while
it is much harder to know if the page's text changed once or twice.
This fact allows us to use a simple estimator for the rate of new
links appearance, which reflects timely trends. Of course, more
sophisticated methods (e.g., using machine learning
\cite{ContentChange}), can be applied here, but it was out of focus of
the current paper.

Moreover, our method actually monitors content sources updates to avoid missing
new links.  In this way, our work is more related to the problem of
creating an efficient RSS feeds reader, which needs to monitor RSS feeds
updates to avoid missing new postings
\cite{FastNewsAlert1,FastNewsAlert2}.
The RSS reader described in these papers learns the general posting pattern
of each RSS feed to create an efficient scheduling algorithm that
optimizes the retrieval of RSS feeds in order to provide timely
content to users.

This RSS reader uses a short description from the RSS feed, when
presenting an article to users, and there is thus no need for it to
crawl these articles.  It is not exactly our case as we need to crawl and
index newly discovered pages to allow users to access them via the
search engine.  Thus, although RSS monitoring policies are somehow
similar to the problem of finding ephemeral new pages on content
sources, one cannot use such policies out of box for our task.  The main
reason is that we need to spend significant amount of resources for
crawling new pages and, if we want to do this in an efficient way, then
we need to change the recrawl schedule to take this fact into account.
However, RSS feeds themselves can be used in our approach
as content sources.

\paragraph{Discovery policies}

The main idea behind discovery policies is to focus breadth-first
search on high quality content, i.e., to prioritize discovered, but not
yet crawled pages (the crawl frontier) according to some quality
measure. Some approaches are based on the link structure of the Web
graph, e.g., in \cite{Autho} pages with the largest number of incoming
links are crawled first, while pages with largest PageRank are prioritized in \cite{OPIC,
  RankMass}.  In \cite{fetterly}, such approaches were compared in
their impact on Web search effectiveness.  However, as Pandey and Olston
discussed in \cite{Olston2}, the correlation between link-based
importance measures and user interest is weak, and hence they proposed to use
search engine logs of user queries to drive the crawler towards pages with higher potential to be interesting for users. In turn, we follow recent trends
and do not rely on the link structure of the Web, but use clicks from
search engine logs.

Our crawler discovers and crawls new pages, but there is a principal
difference between our approach and previous ones.  Previous
approaches are based on the assumption that the Web is relatively deep
and therefore, starting from some seed pages, a crawler needs to go
deeper, in direction of high-quality pages if possible, to find new
pages.  We instead argue that one can find most ephemeral new pages
that are appearing on the Web at a relatively small set of content sources, but
that a crawler needs to frequently recrawl these sources to avoid
missing short living links to new pages.  This observation, on one
hand, simplifies the problem but, one the other hand, introduces new
challenges to find the right balance between crawling new pages and
recrawling content sources.

\paragraph{Holistic crawl ordering}

Usually, papers about crawling focus either on discovery of new pages
or on refreshing already known pages, but the important question of
how to divide limited resources between refreshing and discovery
policies is usually underestimated. Some authors proposed to give a
fixed quota to each policy \cite{Autho, Sitemaps}. However, as it
follows, e.g., from our analysis (see Section \ref{Theory}), such fixed
quotas can be far from optimal.  In contrast, our optimization
framework simultaneously deals with refreshing and discovery and can
thus find an optimal way to share resources.  Moreover, the problem of
making a holistic crawl ordering, i.e., to unify different policies
into a unified strategy, was proposed by Oslton and Najork as a future
direction in their extensive survey on Web crawling \cite{WebCrawling}
and we tried to make a step forward in this direction.

\section{Conclusion}\label{Conclusion}

To the best of our knowledge, the problem of timely and holistic crawling of ephemeral new
content is novel. In this paper, we introduce the notion of ephemeral
new pages, i.e., pages that exhibit the user interest pattern shown on
Figure~\ref{fig:yabar}, and emphasize the importance of this problem
by showing that a significant fraction of the new pages that are
appearing on the Web are ephemeral.

We formalized this problem by proposing to optimize a new quality
metric, which measures the ability of an algorithm to solve this
specific problem. We showed that most of the ephemeral new content can
be found at a relatively small set of content sources and suggested
an algorithm for finding such a set. Then, we proposed a practical
algorithm, which periodically recrawls content sources and crawls
newly created pages linked from them as a solution of this problem. This
algorithm estimates the quality of content sources using user
feedback.

Finally, we compared this algorithm with other crawling strategies
on real-world data and demonstrated that the suggested algorithm shows the
best results according to our metric.  Our theoretical and
experimental analysis aims at giving a better insight into the current challenges
in crawling the Web.



In this paper, we predict the expected profit of a new page using two
features: the time when this page was discovered by a crawler, and
the content source where a link to this page was found. The natural next step, which we leave for future work, is to predict this profit using more features, e.g., give a higher
priority to pages having an anchor text related to the current trends in
user queries like in \cite{Olston1}, or to the pages with more incoming
links. Also, URL tokens and its hyperlink context (anchor text,
surrounding text, etc.) may be useful for such prediction. This
will help to prioritize new pages with seemingly higher quality found on the same content source at the same time.

\end{document}